\documentclass{article}
\usepackage{amssymb}
\usepackage[utf8]{inputenc}
\usepackage{caption}
\usepackage{algorithm}
\usepackage{algpseudocode}
\usepackage{amsmath}
\usepackage{graphicx}
\usepackage{enumerate}
\usepackage[inline,shortlabels]{enumitem}
\usepackage{csquotes}
\usepackage{ulem}
\usepackage{booktabs}

\DeclareMathOperator*{\argmax}{arg\,max}

\usepackage{arxiv}

\usepackage[utf8]{inputenc} 
\usepackage[T1]{fontenc}    
\usepackage{hyperref}       

\usepackage{cleveref}
\usepackage{url}            
\usepackage{booktabs}       
\usepackage{amsfonts}       
\usepackage{nicefrac}       
\usepackage{microtype}      
\usepackage{lipsum}
\usepackage{graphicx}
\graphicspath{ {./images/} }

\title{Smart caching in a Data Lake for High Energy Physics analysis}

\author{
  Tommaso Tedeschi \\
  Department of Physics and Geology\\
  University of Perugia\\
  and\\ 
  INFN - Sezione di Perugia\\
  Perugia, Italy, 06123 \\
  \texttt{tommaso.tedeschi@pg.infn.it} \\
   \And
 Marco Baioletti \\
  Department of Mathematics and Computer Science\\
  University of Perugia\\
  Perugia, Italy, 06123 \\
  \texttt{marco.baioletti@unipg.it} \\
  \And
 Diego Ciangottini \\
INFN - Sezione di Perugia\\
  Perugia, Italy, 06123 \\
  \texttt{diego.ciangottinii@pg.infn.it}
    \And
 Valentina Poggioni \\
  Department of Mathematics and Computer Science\\
  University of Perugia\\
  Perugia, Italy, 06123 \\
  \texttt{valentina.poggioni@unipg.it}
    \And
 Daniele Spiga \\
INFN - Sezione di Perugia\\
  Perugia, Italy, 06123 \\
  \texttt{daniele.spiga@pg.infn.it}
    \And
 Loriano Storchi \\
  Department of Pharmacy\\
  University "G. D'Annunzio" of Chieti-Pescara\\
    Chieti, Italy, 06100 \\
  and\\
  INFN - Sezione di Perugia\\
  Perugia, Italy, 06123 \\
  \texttt{loriano@storchi.org}
  \And
  Mirco Tracolli \\
  INFN - Sezione di Perugia\\
  Perugia, Italy, 06123 \\
  \texttt{m.tracolli@gmail.com}
}

\begin{document}
\maketitle

\begin{abstract}

The continuous growth of data production in almost all scientific areas raises new problems in data access and management, especially in a scenario where the end-users, as well as the resources that they can access, are worldwide distributed. This work is focused on the data caching management in a Data Lake infrastructure in the context of the High Energy Physics field. We are proposing an autonomous method, based on Reinforcement Learning techniques, to improve the user experience and to contain the maintenance costs of the infrastructure.

\end{abstract}

\section{Introduction}

\begin{figure}[htb]
    \begin{center}
        \includegraphics[width=0.75\textwidth]{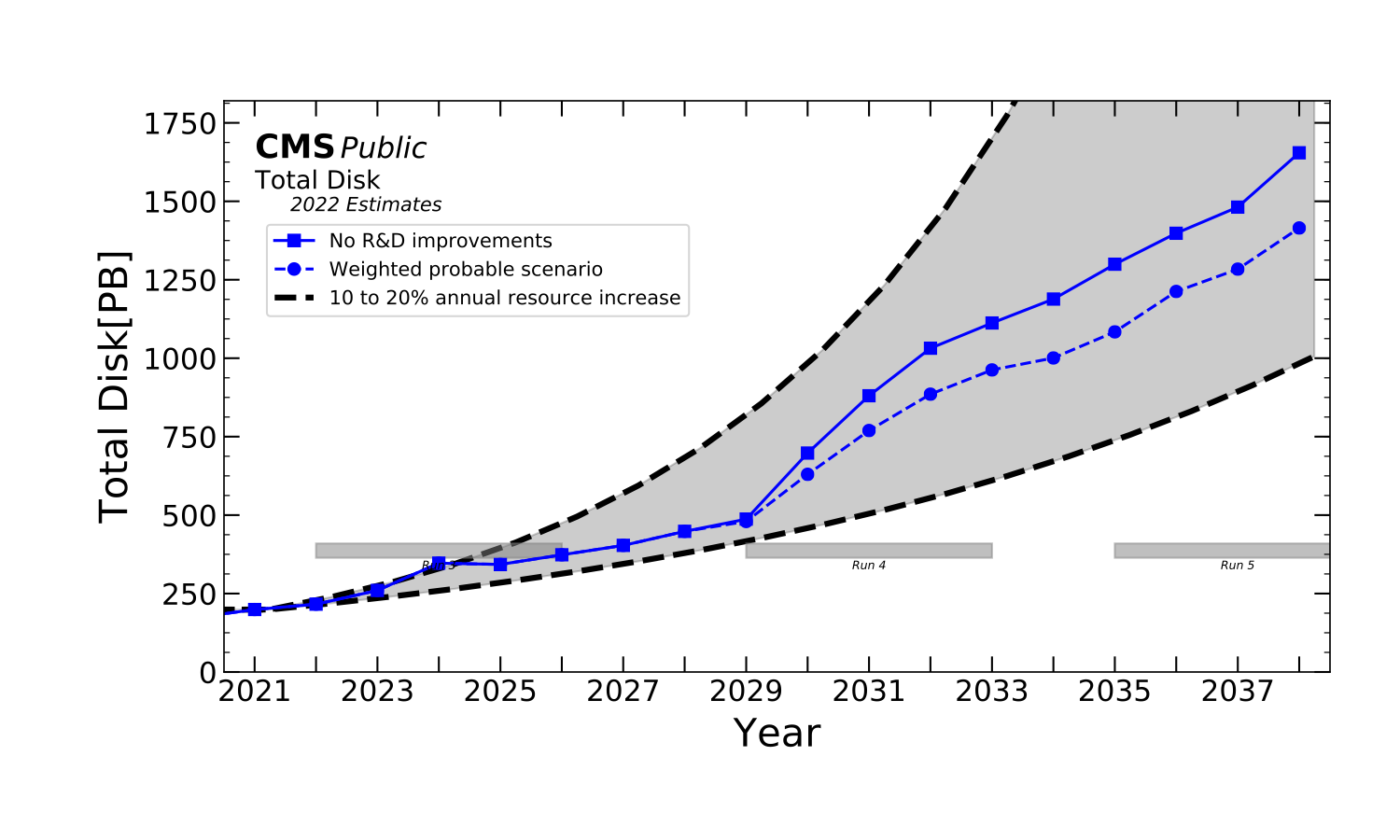}
        \includegraphics[width=0.75\textwidth]{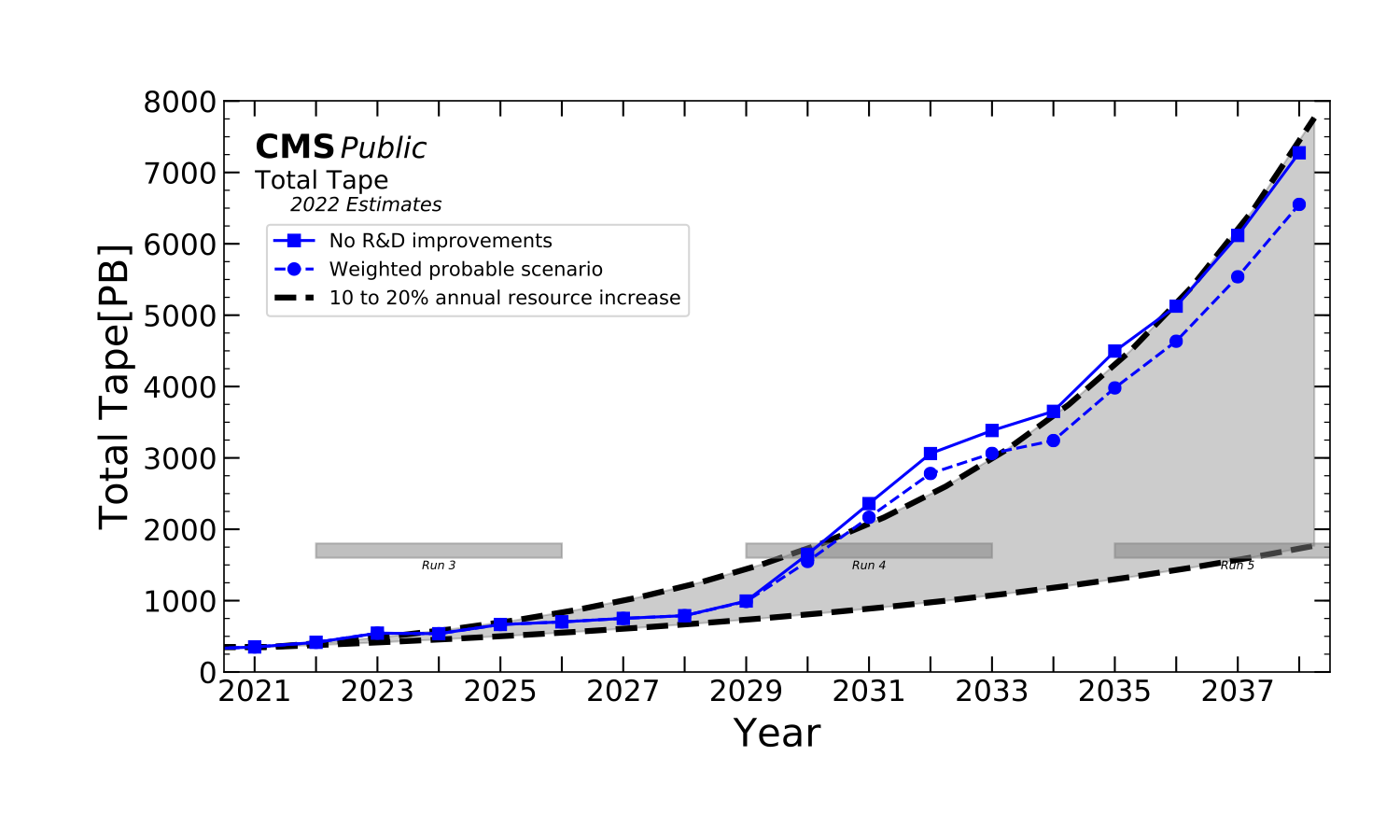}
    \end{center}
    \caption{2022 projections of the increase of data storage at CMS (both for disk and tape), taken from ~\cite{offcompupdate}}
    \label{fig:increaseStorageProjection}
\end{figure}

The Large Hadron Collider (LHC) \cite{Pettersson:291782} at CERN (the European Organization for Nuclear Research) is the world's largest and most powerful particle accelerator. The particle beams inside the LHC are made to collide at four locations around the accelerator ring, corresponding to the positions of four particle detectors: ATLAS~\cite{aad2008atlas}, CMS~\cite{collaboration2008cms}, ALICE~\cite{aamodt2008alice}, and LHCb~\cite{alves2008lhcb}.
A critical challenge at LHC is the next generation of the accelerator expected for 2029, when the named High-Luminosity Large Hadron Collider (HL-LHC) will be fully operative: the upgraded machine will reach an instantaneous luminosity of at least $5 \times 10^{34}$  $\mathrm{cm^{-2}s^{-1}}$ and a center of mass energy of $14$ TeV (with respect to $2 \times 10^{34}$  $\mathrm{cm^{-2}s^{-1}}$ and $13.6$ TeV of current taking period). As a result, data will be produced at higher rates, with a greater event complexity. Consequently, computing resources and storage requests from LHC experiments will increase: as an example, for the disk storage, the future costs are estimated to ultimately be around nine times higher than today (\Cref{fig:increaseStorageProjection}). 

With such expectations, in particular considering that the system will start to manage Exabytes (instead of Petabytes) of data, it becomes clear that software and computing of the experiments, as well as the model adopted, must be reviewed and improved through an intensive R\&D activity, which represents the key to lower future requests (moving from the solid line to the dashed one in \Cref{fig:increaseStorageProjection})

Recently, many architectural, organizational, and technical changes have been investigated to address the challenge of introducing a new data management model and one of the most promising prototypes is the Worldwide LHC Computing Grid (WLCG) Data Lake model~\cite{bird2019architecture,kadochnikov2018wlcg}, a storage service of geographically distributed data centers connected by a low-latency network.
In this model, from the infrastructural perspective, caching systems are used to mitigate latency and to serve better popular data. 

The final goal of the present work is to provide an efficient Reinforcement Learning (RL)-based caching system  which could be used by the WLCG collaboration, and more specifically for the CMS experiment~\cite{collaboration2008cms}, which is one of the biggest experiments at CERN and deploys its data collections, simulation, and analysis activities on a distributed computing infrastructure involving more than 60 sites worldwide.

The work is organized as follows. In Section \ref{sec:dataLakeAtWLCG} a brief description of the Data Lake architecture is given. Then, an introduction to the background concepts needed to understand the project is provided in Section \ref{sec:background}.
In Section \ref{sec:Algorithms} the proposed approaches are presented and discussed in detail, and comparisons with other solutions are described. 
Section \ref{sec:experimentalEnvironment} contains
a description of the experimental environment, while
the experimental results are described and commented in Section \ref{sec:experimentalResults}.
Finally, Section \ref{sec:conclusions} provides some conclusive remarks.

\section{Data Lake at WLCG}
\label{sec:dataLakeAtWLCG}

The Data Lake architecture is designed with the intent of decreasing the cost and to leverage the economy of scale. The WLCG community calculated that storage and computing resources necessary in the future with the scheduled upgrades will dramatically increase despite any current technology evolution \cite{kadochnikov2018wlcg}.

The proposed model is meant to
reduce storage costs, abstract the data layer and manage better the current facilities. It is named \textit{"Data Lake straw model"} and it is a declination of the canonical Data Lake that involves not only the information management~\cite{bird2019architecture}. 

It can be seen as a group of data and compute centers with no defined borders by construction. The world is divided into several Data Lakes that are associated with a specific network latency. The internal Data Lake configuration may vary on the scope of the community and experiment needs.

Still, the main aspect is that data access needs a fast response time, whose order of magnitude can vary a lot depending on the specific type of processing (i.e., access data to use in a specific flow), as well as processing patterns. Therefore, besides the specific optimization of algorithms and software processes, there are several aspects involving the infrastructure level of the analysis environment that could be enhanced. Moreover, the optimization of the access layer becomes more and more important when dealing with a geographically distributed environment, where data must be retrieved from remote servers of a Data Lake~\cite{dataLakeDixon}.

To this purpose, a Data Lake hosts a distributed analysis working set of data and several caches used to reduce the impact of the latency as well as the network load. The data can be relocated from one Data Lake to another, and the most popular datasets may be hosted in more than one Data Lake.

The environment components included in the Data Lake model are:

\begin{itemize}
    \item Archive Center (AC): responsible for archive custodial data, the source of information. It should use non-performant storage like tape drivers;
    \item Data and Compute Center (DCC): focused on disk storage faster than AC (mechanical hard disks) but with also computing power, it is used to increase the quality of service (QoS) for analysis tasks;
    \item Compute Center with Cache (CCC): it is a center focused on calculus without memory to store the data but with a fast cache to serve the analysis jobs;
    \item Compute Center with Direct Access (CCDA): a poorer version of a Compute Center having also a lower volume of the cache. It relies especially on the network to access data. It has no disk space and consumes computing jobs taking data from either a CCC or a DCC.
\end{itemize}
It is clear that the role of the cache becomes the key to effective and efficient data access, while saving the storage needs of the experiment.

\begin{figure}[htbp]
    \centerline{\includegraphics[width=0.9\textwidth]{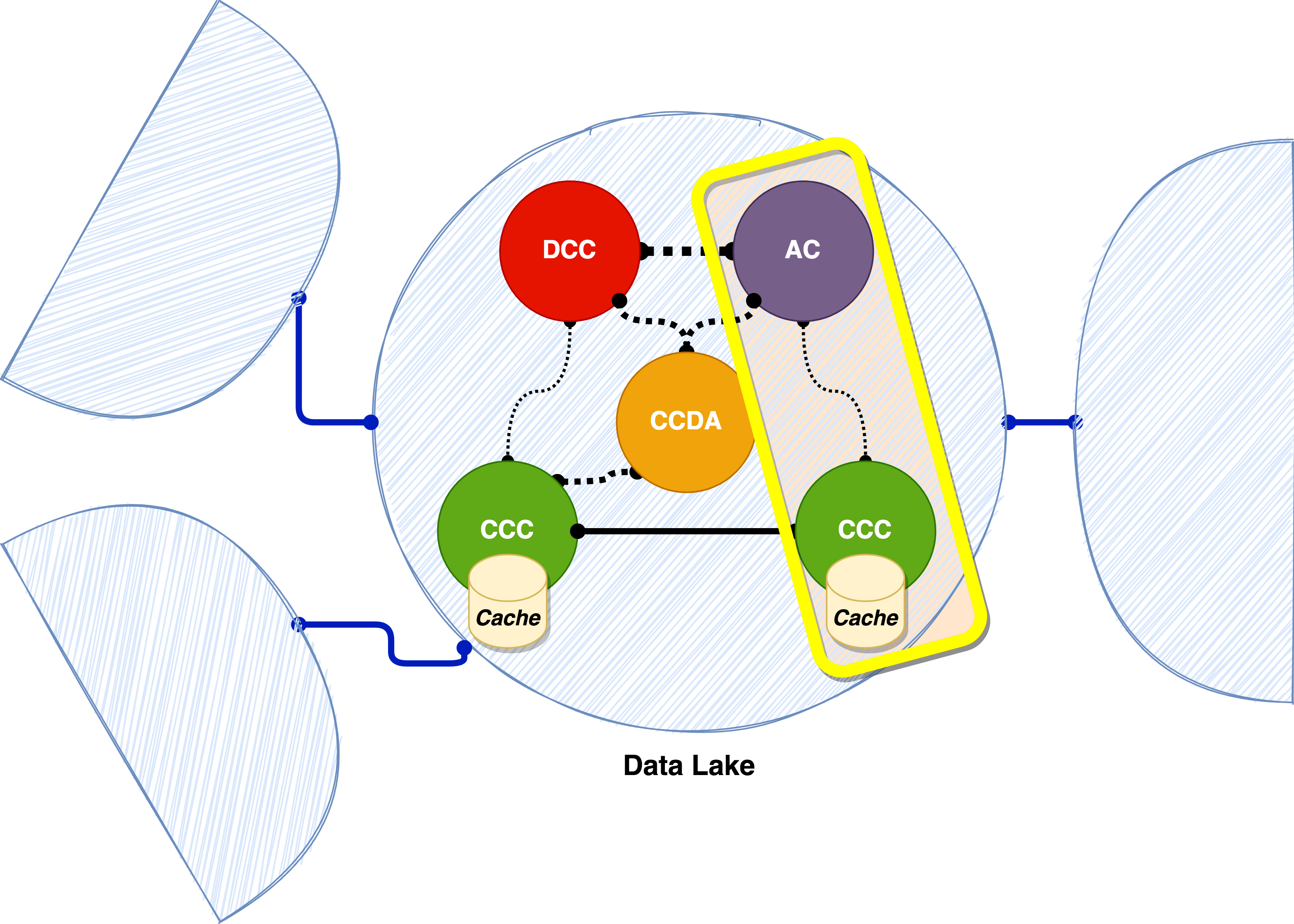}}
    \caption{Illustration of environment implementation with a focus on the components selected to solve the problem}
    \label{img:problemFocus}
\end{figure}

In this work we focus on the CCC component reading from an AC, evaluating and optimizing the performances of the CCC cache system, as depicted in in~\Cref{img:problemFocus}.

In terms of caching data management our main goal is to solve a problem that has many affinities with a Content Delivery Network, and with the web content caching (especially when video file streaming is considered~\cite{adhikari2012unreeling}). However, this project specifically targets the  High Energy Physics research community, 
that need to optimize the data access in the Data Lake environment while 
making the system more autonomous to avoid the human intervention as much as possible. For these reasons we chose a Reinforcement Learning (RL) 
approach\cite{sutton2018reinforcement}, that learns to interact directly with the environment, self-adapting to new situations in the context of Data Lakes, despite the domain of the data or the current topology of the network.

To summarize, the model we are considering is made of three basic components: the main storage system (i.e., where the files reside), a cache that serves the requests, and a client that requires the data. The main goal of the caching system is clearly to resolve all the client's requests and serve the files from the cache. This simplified model allows testing different policies to control the request flow. 

\section{Background on Reinforcement Learning}
\label{sec:background}

The approach used in this work is Reinforcement Learning (RL), which is one of the most important methods in Machine Learning (ML), and aims at training an agent to interact with a particular environment.

RL differs from the other types of ML because it puts the learner in a situation of trial and error, where the consequences of its actions have an impact on the environment and also on the problem's goal. Furthermore, the agent is punished or rewarded on the basis of its behavior, with the idea that, in the future, it will prefer optimal actions and forego unwanted behaviors. 
As a consequence, RL is focused on goal-directed learning from interaction. For this reason, it differs from Supervised Learning because it does not use a set of labeled examples provided by a knowledgeable external supervisor. 

One challenging aspect of RL algorithm is the trade-off between exploration (i.e., trying new actions) and exploitation (i.e., applying what was learned). The balance between them remains an unresolved problem and one of the most delicate parameters to set.

RL has fruitful interactions with other engineering and scientific disciplines and can fit a variety of problems (e.g.~\cite{mnih2013playing}), also because it could be an independent component of a larger behaving system.
For further details, see \cite{sutton2018reinforcement}.

\subsection{Environment Description}
\label{sec:rlenvironment}

\begin{figure}
\centering
\includegraphics[width=.5\textwidth]{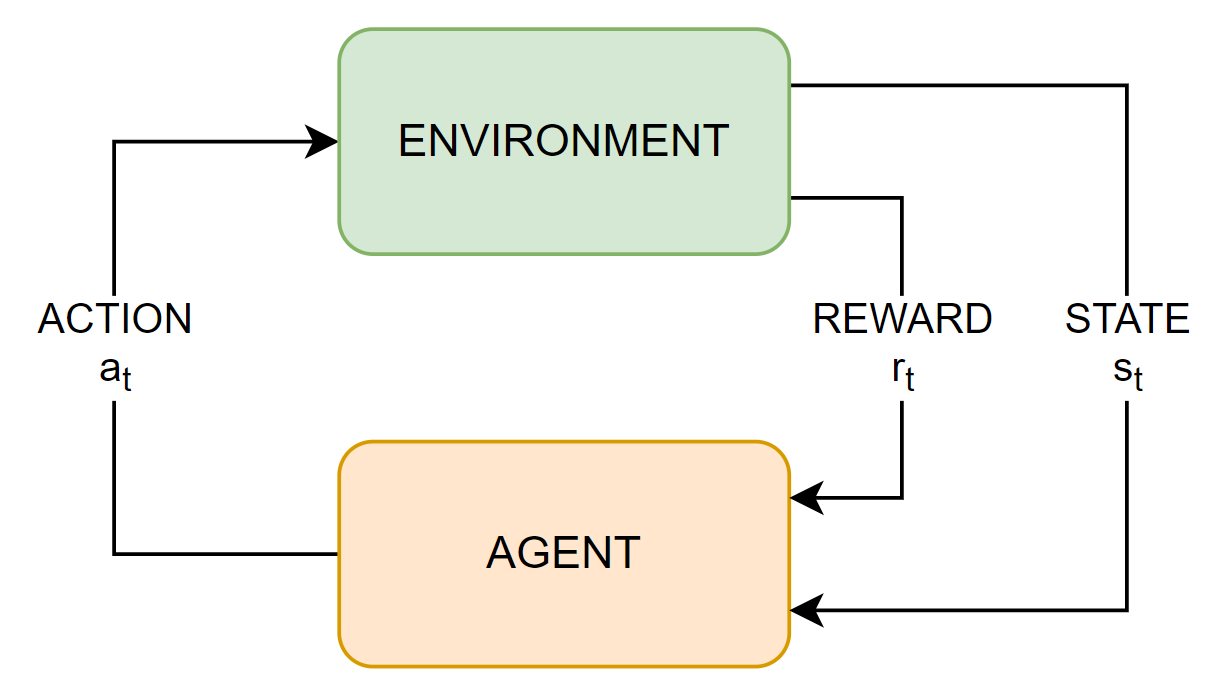}
\caption{Reinforcement Learning schema.}
\label{fig:RLstep}
\end{figure}

The agent trained in RL continuously interacts with the environment as shown in a schematic way in \cref{fig:RLstep}. At each time step,
the agent observes the environment, obtaining a state $s$, and chooses a certain action $a$ to execute, according to a given policy $\pi$. As a consequence, it receives a reward (which can be negative, i.e. a punishment) $r$ from the environment. 
The ultimate goal is to maximize its cumulative reward, the so-called \textit{return}, hence finding the optimal policy $\pi^*$, which maximizes the expected \textit{return} when the agent acts correctly.

The Optimal Action-Value Function $Q^*(s,a)$ is the function that computes the expected \textit{return} if, starting from $s$, the action $a$ is executed applying the policy $\pi^*$. Hence, the optimal action is selected as:
\begin{equation}
    a^*(s)=\argmax_a{Q^*(s,a)}
    \label{eq:optimalaction}
\end{equation}

Moreover, the Optimal Action-Value Function $Q^*(s,a)$ obeys to the self-consistency Bellman equation:
\begin{equation}
Q^*(s,a) = \displaystyle \mathop{\mathbb{E}}_{s' \sim P(\cdot \| s,a)}[r(s,a) + \gamma \max_{a'}Q^*(s',a')]
    \label{eq:bellman}
\end{equation}

where $s'$ identifies the next state (sampled from the distribution $P(\cdot \| s,a)$ governing all environmental transitions) and $\gamma \in [0,1]$ is the so-called \textit{discount factor}.

\subsection{Q-learning and Deep Q-learning}
\label{sec:qlearning}

Q-Learning is one of the best known RL methods: in its simplest form, the agent tries to learn the $Q^*(s,a)$ function by acting $\epsilon$-greedily, i.e. by selecting a random action $a$ with probability $\epsilon$ (that decays over time), otherwise by selecting an action $a$ according to \cref{eq:greedy}.

\begin{equation}
    a(s)=\argmax_{a'}{Q(s,a')}
    \label{eq:greedy}
\end{equation}

The first behavior allows the exploration of all possible actions, whereas the second one allows the exploitation of the knowledge gained by the agent.

Learned Q-values are stored in a tabular form for each pair $(s,a)$. The particular Q-value is updated at each step accordingly to  \cref{eq:Qlearning}

\begin{equation}
    Q'(s,a) \longleftarrow  Q(s,a) + \alpha(r_t + \gamma \max_{a'}Q(s_{t+1},a') - Q(s_t,a_t)) .
    \label{eq:Qlearning}
\end{equation}

The memory and computation required for the Q-value algorithm would be too high for real-world scale problems thus, in several applications, Deep Neural Networks (DNN), which approximates the Q-Learning functions, are used (Deep RL).

In the present work, following the approach proposed by Mnih et al. \cite{mnih2015human}, the Q-value function $Q(s,a)$ is approximated by a DNN, while the objective function is still based on the Bellman Equation in \cref{eq:bellman}.  Moreover an experience replay buffer, as well as a target network, are used to guarantee a stable training. This learning algorithm is called Deep Q-Network (DQN).

\section{Algorithms}
\label{sec:Algorithms}

\begin{figure}[htbp]
\centerline{\includegraphics[width=0.9\textwidth]{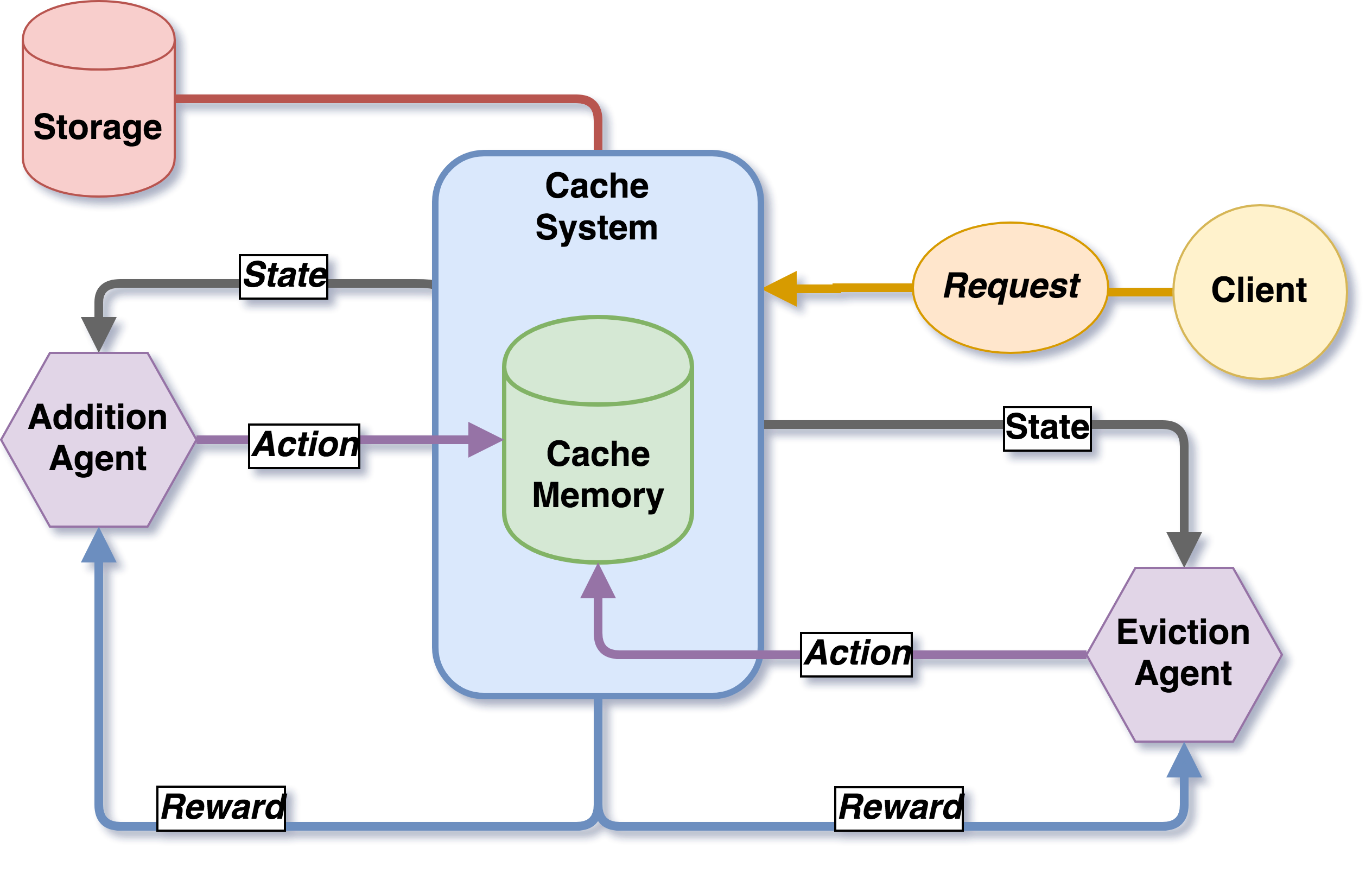}}
\caption{Reinforcement Learning schema of the double agent approach, where the AI chooses both the addition and eviction of a file into the cache memory}
\label{img:scdl2Schema}
\end{figure}

In the following, we will describe the three different caching algorithms we have implemented and tested. As for the inner design of the framework, all the three algorithms are based on the diagram in \Cref{img:scdl2Schema} . As depicted in the schema, the cache system can interact with two agents: the first one is trained to decide whether to add a file to the cache or not, while the latter one decides which files have to be removed from the cache memory.

We treated the problem with an incremental approach with respect to algorithm functionality, and we developed and tested three different caching algorithms. 
The first algorithm, based on the Q-Learning method, uses a single agent that is in charge of choosing the files to be added to the cache and employing the LFU (Least Frequently Used) policy~\cite{podlipnig2003survey}
for choosing the files to be removed. 
The second, based on the Q-Learning method, implements two agents to decide which files to add to the cache and which categories of files to delete, respectively.
Finally, the third, using the DQN method, implements two agents in which also the deletions are defined with respect to single files. 
 The two double-agent approaches are schematized in Figure \ref{img:scdl2Schema}, where both the interactions of the Eviction and Addition agents are represented. 

\subsection{Related works}
\label{sec:relatedWorks}

The most used strategy to manage caches is a "Write Everything" approach associated to a Least Recently Used (LRU) or to a Least Frequently Used (LFU) eviction policy ~\cite{podlipnig2003survey}. 
They  can be effective in most of the cases, but they cannot deal  with  content popularity and network topologies changing over the time. Hence, recent efforts have gradually shifted toward developing learning and optimization-based approaches, and several ML techniques have been proposed to improve file caching and, in general, better content management. 

L. Lei et al.~\cite{lei2017deep} propose to train a DNN in advance to better manage the real-time scheduling of cache content into a heterogeneous network. 
In \cite{narayanan2018deepcache} a Deep Recurrent Neural Network is applied to predict the cache accesses and to make a better caching decision, but this work has been applied just to cache and synthetic dataset whose sizes are far from the Data Lake volumes.
Another example of a prediction approach is presented in~\cite{lykouris2018competitive}, where predictions are used to optimize the eviction of a cache with a fixed size. 
While, in~\cite{herodotou2019autocache}, an attempt to automate the caching management of a distributed data cluster using the Gradient Boosting Tree is presented. 

It is evident that the environment is a critical aspect that has to be taken into account when we talk about caching management and, due to its variability, a more flexible and autonomous solution that can adapt itself is needed.
To meet this need, techniques based on RL approach have been recently proposed. 
In \cite{sadeghi2019deep} a Deep RL approach is used to cache the highly popular contents across distributed caching entities in the context of CDN. However, even if the system allows an online adaptation, the experiment uses a few files that have to be placed optimally in a hierarchical caching system.
There are also Deep RL approaches, like the Wolpertinger architecture~\cite{dulac2015deep} used by C. Zhong et al.~\cite{zhong2018deep}, that try to optimize the cache hit rate. But, in that case, the authors assume that all the files in the cache have the same size, and this is not always the case in High Energy Physics context.

Thus, the problems solved by the cited works are not fully comparable in size and needs with respect to the ones that we are targeting in our project, where there are a much larger number of files to manage and a huge amount of requests per day to satisfy.
Moreover, the field of application is different and very specific. The High Energy Physics context has a data access pattern that cannot be always directly compared with respect to other use cases as we are dealing with an heterogeneous community of users chaotically producing files of different size and structure. Furthermore, there is a real necessity to meet the future requirements with the current budget constraints, otherwise the user experience will be drastically compromised.

For these motivations, we propose three different RL-based algorithms to tackle the cache decisions in terms of file eviction and addition, similarly to what is done in \cite{alabed}. In the next sections, we will describe the algorithms in detail.
In the first two algorithms
a similar mechanism to the one used by the caching system accordingly the WLCG
Model (~\Cref{sec:dataLakeAtWLCG})
 is used to  prevent the cache memory to become too full
or too empty. The mechanism is based on a high (i.e. $W_{high}$) and a low (i.e. $W_{low}$) watermark. When the $W_{high}$ memory occupation is reached a file deletion process starts and continues until $W_{low}$ memory occupation is reached.
The two watermarks are set to 95\% and 75\% of the cache size, respectively for $W_{high}$ and $W_{low}$. The last algorithm, i.e. DQN QCache, only uses the $W_{high}$ watermark.

\subsection{SCDL}
\label{sec:SCDL}

The Smart Cache for Data Lake (SCDL) \cite{scdl_code} algorithm is the first approach proposed. It is based on the Q-Learning method, and 
uses information taken from both the files to be managed and the state of the cache. Nevertheless, SCDL is not a full Reinforcement Learning technique, but rather uses a variant: the associative search task, known also as \textit{contextual bandits} \cite{SCDL}.

The most important characteristic is that it implements only an Addition Agent. Thus, an agent that can only discriminate either to write or not write a file following a client request. The deletion of files is performed according to the LRU policy using the watermark mechanism. 

Each time the user requests a file $f$ to the cache, the state $s$ used by the agent is computed in terms of a set of statistics related to $f$, collected during the environment lifetime: the file size $s_f$, the number of requests of the given file $n_f$, and the delta time $\Delta t_f$, that has passed since the last request of $f$. It is assumed that the environment uses a discrete internal time that is incremented at each request.
The statistic history traces $7$ days of file's requests, and it is deleted if the file is no longer present in the cache memory. 

Since the number of states must be finite due to the Q-Learning algorithm, the file features are discretized in a finite number of classes using a simple binning technique with ranges. 
Hence, the states for the addition agent are defined as:

\begin{equation}
    S_a=b(s_f, n_f, \Delta t_f) 
    \label{ref:SCDLAgentState}
\end{equation}

where $b$ is the function that returns a 3-ple of the corresponding class in $s_f$, $n_f$ and $\Delta t_f$.

For each possible state, there are two possible actions: \textit{Store} and \textit{NotStore}. Because the state does not change after the agent decision, the environment faced by the algorithm is similar to the contextual bandit case.
Indeed, the next state $s'$ (Formula~\Cref{eq:Qlearning}) is the same as the input state $s$. 

The decision taken by the agent is rewarded in a delayed way: each action $a$ taken on a state $s$ is memorized and will be rewarded later, when $s$ occurs again. 
We assign a positive reward $r$ equals to the \textit{size} of $f$, if the cache constraints (defined a priori) are satisfied. Conversely, a negative reward $-$\textit{size} is assigned.
A pseudocode of the whole algorithm can be found in~\Cref{code:scdl}.

\begin{algorithm}
\caption{Smart Cache for Data Lake (SCDL) algorithm pseudocode}
\label{code:scdl}
\begin{algorithmic}
\Function{SCDL}{request}
\State file $\gets$ request.filename
\State update the statistics with request
\State hit $\gets$ cache\_search(file)
\If{\textbf{not} hit}
    \If{random$<\epsilon$}
        \State action$\gets$random\_action(state)
    \Else
        \State action$\gets$best\_action(state)
    \EndIf
    \If{action is Store}
        \State cache\_add(file)
    \EndIf
\EndIf
\State delayed\_reward(state)
\EndFunction
\end{algorithmic}
\end{algorithm}

\subsection{SCDL2}
\label{sec:SCDL2}

SCDL2 (Smart Cache for Data Lake 2) \cite{scdl_code} is an evolution of the previous approach that uses two different agents to solve the caching problem: the Addition Agent decides whether a requested file has to be stored, while the Eviction Agent chooses how to free the cache memory removing all the files belonging to a specific file category. While the Addition Agent focuses its decision on the state of the request, the eviction agent decisions depend more on the state of the cache memory.

As shown in \Cref{img:scdl2Schema}, the goal is to modify the policies used by the cache to decide whether to store a file and, in case space is needed, what to evict. A pseudocode of this approach is available in~\Cref{code:scdl2}.

When a file $f$ is requested, the Addition Agent is called in order to decide  whether to store or not the file $f$. 

The Eviction Agent is called only in particular situations: when it is necessary to free space in the cache, at the end of each day, after $k$ iterations (corresponding to the number of requests made to the cache).  In those situations, it chooses which files to remove.

The two agents work with different state spaces: the Addition Agent's state $s$ is quite similar to the SCDL algorithm. In this version, the state is enriched with the cache occupancy percentage $oc$ and cache hit rate $hr$. As a consequence, states for the Addition Agent are defined as:

\begin{equation}
    S_a=b(s_f, n_f, \Delta t_f, oc, hr) 
    \label{ref:SCDL2AgentState}
\end{equation}

where $b$ is the binning function.

On the other hand, the Eviction Agent composes its state based on the cache memory content. Specifically, the files stored in the cache memory are split into categories subsequently used to choose the set of files to remove. Similarly to the Addition Agent, the Eviction Agent uses $s_f$, $n_f$, $ \Delta t_f$ to associate the file to a specific category $c$, that contains all the files of a specific size $s_c$, that have been requested $n_c$ times and for which $\Delta t_c$ time has passed since the last request. Moreover, for each category, also the amount of space occupied by the category itself, named $oc_{c}$, is considered.

Those features, together with the features characterizing the state of the cache $oc$ and $hr$, are then discretized in a finite number of buckets, and they result in the following state definition:

\begin{equation}
    S_{e} = b_{e}(s_{c}, n_{c}, \Delta t_c, oc_{c}, oc, hr) 
    \label{ref:SCDL2EvictionAgentState}
\end{equation}
where $b_e$ is the function mapping $s_c$, $n_c$, $\Delta t_c$, $oc_{c}$, $oc$ and $hr$ to the corresponding classes.

The results of the agents' decisions are stored into two different Q-tables, where all the actions are evaluated for each possible state: \textit{additionTable} and \textit{evictionTable}.

The action space for the Addition Agent is composed of two possibilities: \textit{Store} and \textit{NotStore}.

The action space for the Eviction Agent contains five possibilities: \textit{NotDelete}, \textit{DeleteAll}, \textit{DeleteHalf}, \textit{DeleteQuarter} and \textit{DeleteOne}, that delete respectively no files, all the files, a random half, a random quarter or a single random file belonging to the category. These methods identify how a selected category has to be managed. The choice of considering a finite number of actions for each specific category, instead of having a different delete action for each file stored in the cache, reduces the agent search space.

Since the decision of storing a file $f$ affects the cache composition, and its actual contribution cannot be determined immediately, we decided to use a delayed reward approach.
Therefore, after each file request, we store the action chosen by the agent. Then, later in time, the agent will evaluate that decision with a positive or a negative reward depending on specific rules.

For the Addition Agent, we assign a positive reward of $r=+1$ to all \textit{Store} actions that allowed a later-requested file to be in memory. The action takes an extra $+1$ if the situation passed from a miss to a hit with that action. 

Similarly, the agent is penalized with a reward $r=-1$ if the file was not in memory, and it chose the \textit{NotStore} action. In the latter case, if the file passed from hit to miss, there is an extra malus of $-1$. 

For the Eviction Agent the rules are very similar, but with the file category as target. In details, a positive reward $r=+1$ is assigned to the action \textit{NotDelete} if the file is found in the cache at the request. Moreover, there is an extra bonus of $+1$ if the cache occupation is not increased in the current request iteration. 

Conversely, a negative reward $r=-1$ is assigned to the action which deleted the file  when a file of a specific category is not found, and an additive malus of $-1$ is given if the file passed from hit to miss.

To summarize, the environment chooses to penalize those actions that cause the cache to perform more work, such as writing new files and removing files to free space. Thus, the agent tries to avoid non-useful operations, and to minimize the cache actions. 

\begin{algorithm}
\caption{Smart Cache for Data Lake 2 (SCDL2) algorithm pseudocode}
\label{code:scdl2}
\begin{algorithmic}
\Function{SCDL2}{request}
\State file $\gets$ request.filename
\State update the statistics with request
\State hit $\gets$ cache\_search(file)
\If{\textbf{not} hit}
    \If{random$<\epsilon$}
        \State action$\gets$random\_action\_from\_addition\_agent(state)
    \Else
        \State action$\gets$best\_action\_from\_addition\_agent(state)
    \EndIf
    \If{action is Store}
        \State cache\_add(file)
    \EndIf
\EndIf
\If{trigger for eviction agent}
    \State call\_eviction\_agent(request)
\EndIf
\State delayed\_rewards()
\EndFunction
\end{algorithmic}
\end{algorithm}

\subsection{DQN QCache}
\label{sec:DQN}
The DQN QCache approach is still based on the two agents depicted in \Cref{img:scdl2Schema} but they are implemented with the DQN approach \cite{dql_code}.

DQN \cite{mnih2015human} basically extends the Q-learning approach by approximating the action-value function with a DNN (\Cref{code:dqcache}). One of the main advantages of this approach is that the input state values are continuous, as compared to the discretization step required by the standard Q-Learning, allowing for a higher granularity. 

Each file request and each cached file, respectively, correspond to a specific input state $S$. Similarly to SCDL and SCDL2 approaches, when a file request arrives, the Addition Agent chooses whether that file has to be cached or not. Whereas, differently from those two approaches, in the DQN QCache approach the Eviction Agent is iteratively applied on every cached file, choosing if it has to be evicted or kept. The latter action is performed every $k$ requests, or when the high watermark $W_{high}$ is reached.

Both agents use the same information about the files. 
A state associated to a requested file, or to a cached one, is represented as 
\begin{equation}
    S=(s_f,n_f,\delta t_f, d_f, oc, hr)
    \label{ref:DQNAgentState}
\end{equation}
where $s_f$ is the file size, $n_f$ is the file frequency, $\Delta t_f$ is the time passed since the last request of the file, $d_f$ is the data type, $oc$ is the cache occupancy percentage, and $hr$ is the cache hit rate. 

Every $N$ requests, the algorithm looks for actions for which were taken at least $h_{window}$ requests ago. Thus, when an elapsed time window is found,  the reward $R$ is computed and the next state $S'$ is defined, starting from $S$, by increasing the frequency by one and using current cache occupancy and hit rate, leaving unchanged all the other information.

 That is:
\begin{equation}
    S'=(s_f,n_f + 1,\delta t_f, d_f, oc(t'), hr(t'))
    \label{ref:DQNAgentStateSprime}
\end{equation}
The 4-tuple $(S,A,R,S')$, where $A$ is the action taken and $R$ is the reward, is stored in the agent's experience replay memory.

The action space for the Addition Agent is composed of the two actions $Store$ and $NotStore$, while the action space for the Eviction Agent consists of the two actions $Keep$ and $NotKeep$. 

The reward, for both agents, is determined by the number of times a certain file is requested by users after a choice has been taken. The algorithm observes $h_{window}$ requests after the action and, depending on the action, computes the reward. Specifically, if the action was $Store$ (for the Addition Agent) or $Keep$ (for the Eviction Agent) the reward is computed as follows:
\begin{equation}
        r = \begin{cases} n_{hit} \cdot s_f, & \mbox{if } n_{hit}>0\\ -s_f, & \mbox{otherwise} \end{cases}
\end{equation}
where $n_{hit}$ is the number of hits for that file in the next $h_{window}$ requests, and $s_f$ is the size of the file. 

If the action was $NotStore$ (for the Addition Agent) or $NotKeep$ (for the Eviction Agent), the reward is:
\begin{equation}
    r = \begin{cases} - n_{miss} \cdot s_f, & \mbox{if } n_{miss}>0\\ s_f, & \mbox{otherwise} \end{cases}
\end{equation}
where $n_{miss}$ is the number of misses for that file in the next $h_{window}$ requests, and $s_f$ is, as usually, the size of the file.

\begin{algorithm}
\caption{\textit{DQN QCache} algorithm pseudocode}
\label{code:dqcache}
\begin{algorithmic}
\Function{DQN QCache}{request}
\State file $\gets$ request.filename
\State update the statistics with request
\State hit $\gets$ cache\_search(file)
\If{\textbf{not} hit}
    \If{random$<\epsilon$ or requests $<$ addition agent warm up counter}
        \State action$\gets$random\_action\_from\_addition\_agent(state)
    \Else
        \State action$\gets$best\_action\_from\_addition\_agent(state)
    \EndIf
    \If{action is Store}
        \State cache\_add(file)
    \EndIf
    \If{\#requests $>$ addition agent warm up counter}
        \State batch$\gets$sample from cache addition memory
        \State train addition agent on batch
    \EndIf
\EndIf
\If{trigger for eviction agent}
    \For{file in cache}
            \If{random$<\epsilon$ or requests $<$ eviction agent warm up counter}
        \State action$\gets$random\_action\_from\_eviction\_agent(state)
    \Else
        \State action$\gets$best\_action\_from\_eviction\_agent(state)
    \EndIf
    \If{\#eviction agent call $>$ eviction agent warm up counter}
        \State batch$\gets$sample from cache eviction memory
        \State train eviction agent on batch
    \EndIf
    \EndFor
\EndIf
\If{trigger look for elapsed time windows}
    \State find\_and\_reward\_elapsed\_actions\_and\_add\_to\_memory()
\EndIf
\EndFunction
\end{algorithmic}
\end{algorithm}

\section{Experimental Environment}
\label{sec:experimentalEnvironment}

Having introduced in details the main characteristics of the three different caching algorithms we implemented, in the following we present the results and the metrics used to compare the different approaches.

\begin{figure}[htb]
    \begin{center}
        \includegraphics[width=0.75\textwidth]{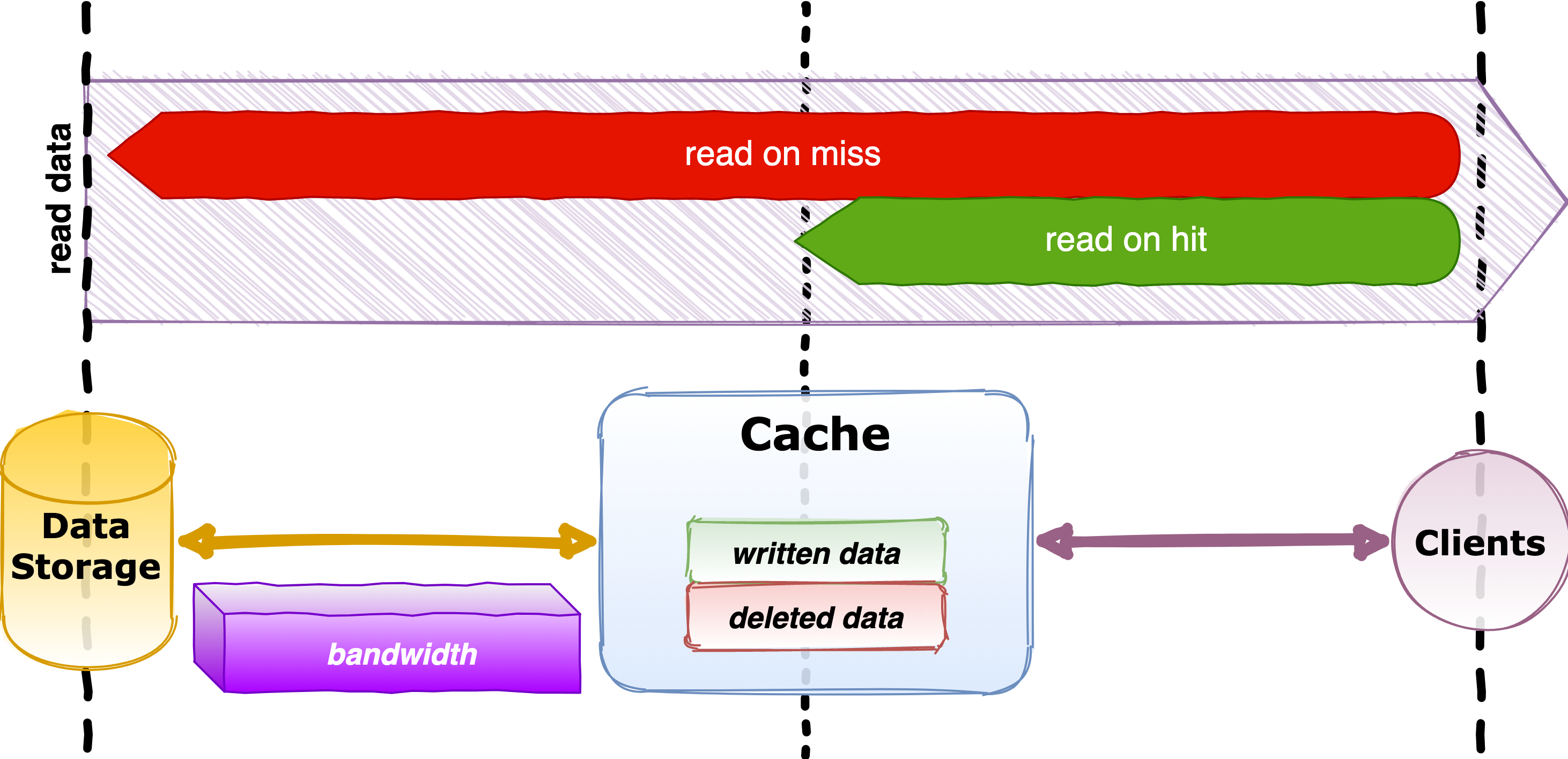}
    \end{center}
    \caption{Simulation environment schema showing the several aspects taken into account and the units measured}
    \label{fig:simulationEnvironmentSchema}
\end{figure}

When the cache decides not to store a file, the latter is served in 
proxy mode, which means that it will fall back on the network.
The \Cref{fig:simulationEnvironmentSchema} shows a schema of the environment and the main statistics collected to 
evaluate the cache behavior. These data allow us to define three evaluation metrics, that will be detailed in following.

Accordingly to the previous description of the environment, and to the schema represented in 
\Cref{fig:simulationEnvironmentSchema}, the data read from the storage is split into two sets: 
\textit{Read on Miss} (i.e.,  data served in the proxy mode because files are not stored in the cache memory), and 
\textit{Read on Hit} (i.e., data served directly from the cache memory).
An ideal cache should be able to keep the \textit{Read on Hit} as high as possible, while keeping the \textit{Read on Miss} as low as possible, aiming to unload as much as possible the main storage server. 

In addition, since the simulator is used to stress the cache decisions, in order to simulate the bandwidth limit a simple threshold for daily requests is used, i.e., if the given limit is exceeded the request is processed as a remote call and, consequently, is counted as a miss (a similar mechanism is used in the real-world caching systems where if a cache is overloaded the requests are \textit{redirected} to other caches).

To conclude, there are several parameters to keep under control for the cache content management 
improvement. It is not trivial to translate the gain obtained by a specific algorithm with 
respect to the final user experience that is strictly related to data access. Of course, the better 
the cached content is managed, the greater is expected to be the end-user experience. However, 
the main goal of the project is to automate and facilitate the management 
of the cache layer for the system maintainers. 

\subsection{Dataset}
\label{sec:datasets}

This work uses information on historical user analysis activities at CMS. In order to get a first feedback on the effectiveness of these approaches we tested caches with different sizes using data coming from the real world. A dataset obtained from historical monitoring data of the CMS experiment analysis jobs related to year 2018 \cite{kuznetsov2016predicting, meoni2018dataset}, filtered for the Italian region, has been used.

\begin{figure}[htbp]
    \begin{center}
        \includegraphics[width=0.9\textwidth]{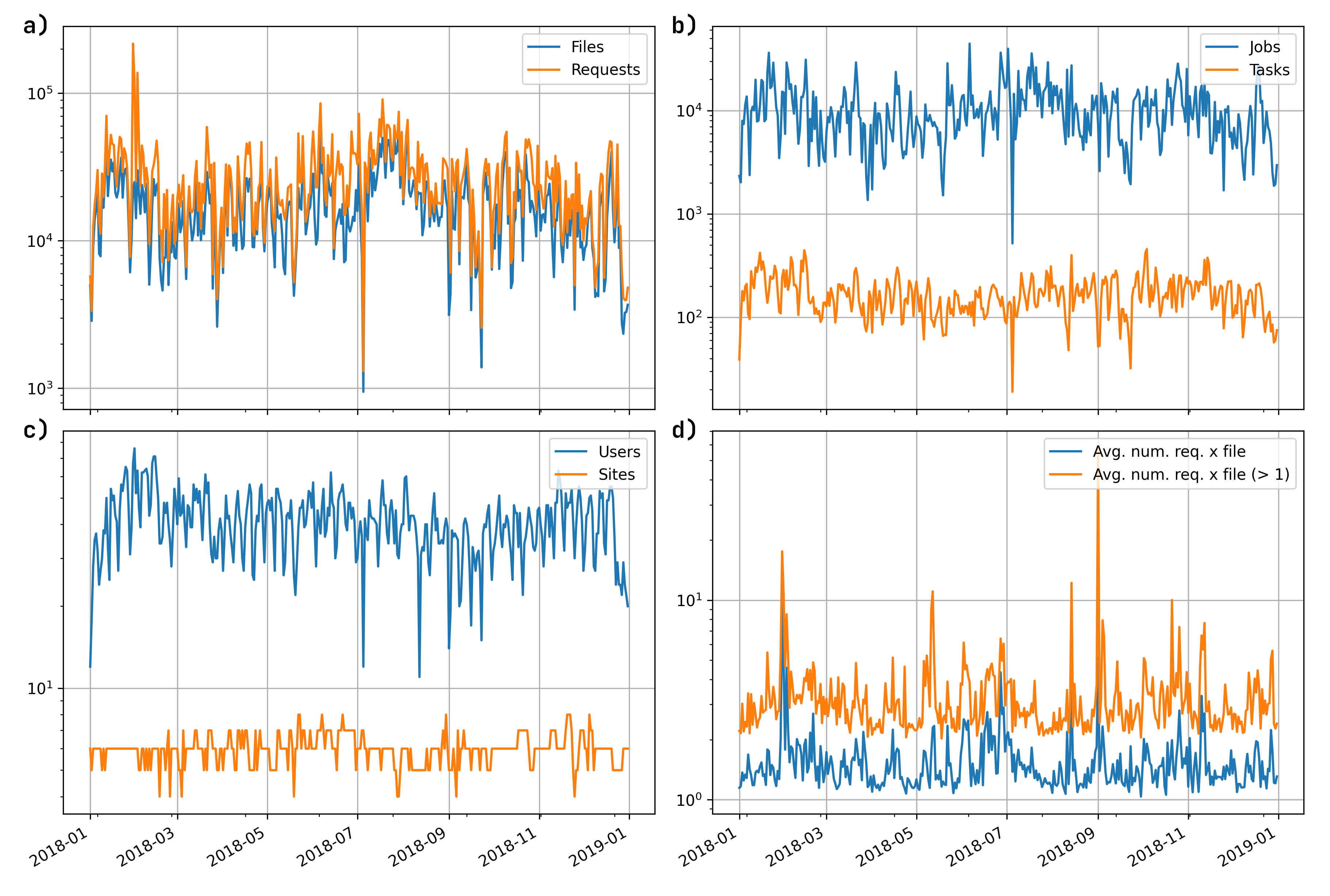}
    \end{center}
    \caption{Data general statistics of Italian requests in 2018 showing the number of files, requests, jobs, tasks, users, and sites during the year}
    \label{fig:stastsGeneralIT}
\end{figure}

To give the reader an overview of the dataset used, in \Cref{fig:stastsGeneralIT} we are reporting four plots. 
Plot $a)$ shows the total number of files and requests per day, where the average is calculated counting only the files requested more than once. 
Plots $b)$ and $c)$ show 
the number of jobs and tasks and the number of users and sites per day, respectively. 
Finally, plot $d)$ shows the daily average number of requests per file as functions of the day of the year.

We can clearly notice that the number of tasks, i.e., group of jobs, is two orders of magnitude lower than the number of jobs that can request several files as input. Furthermore, the number of sites (i.e., the place of the request representing a computing center) is much lower than the number of users. 

More importantly, as reported in plot $d)$, the number of requests per file is low on average (i.e., the average number of requests per file is $\approx 5$ per file, for the files requested more than once), and the number of files requested per day is comparable to the total number of requests per day (the value is greater than $10^4$). To summarize, there are a lot of requests per day but the majority are unique request, thus not an easy scenario for a caching system.

\subsection{Evaluation metrics}
\label{sec:evaluationMetrics}

In order to evaluate and compare the different approaches proposed in \Cref{sec:Algorithms}, we decide to monitor two main aspects of the cache environment (\Cref{fig:simulationEnvironmentSchema}), the \textit{Throughput} (TP) and the \textit{Cost}. The $TP$ is defined as following:
        \begin{equation}
            TP=\frac{RHD}{RHD_{\infty}}
            \label{eq:throughput}
        \end{equation}
where $RHD$ (Read on Hit Data) represents the total amount of data that are read directly from the cache. Since $RHD$ is an absolute quantity that depends on the cache size, we decided to normalize it with respect to the ideal upper bound computed on an infinite cache $RHD_{\infty}$. In this case, the amount of data that can be read directly from the cache corresponds to the total amount
of data that has been written to the cache (i.e., if the cache is infinite we can write any data). 

The \textit{Cost} metric is defined as:
\begin{equation}
    Cost=\frac{WD+DD}{2\cdot WD_{\infty}}
    \label{eq:cost}
\end{equation}
where $WD$ and $DD$ represents the total amounts of written and the deleted data, respectively. 
They are used to measure how much the cache is working in terms of pure cache operations with respect to $WD_{\infty}$, the amount of data  we can write to an ideal infinite cache.

It is important to note that we cannot evaluate our approach considering the sole \textit{hit rate} (i.e. the standard measure used in cache evaluation) because this measure assumes that all the files have the same size.

\begin{figure}[htb]
    \begin{center}
    \includegraphics[width=0.9\textwidth]{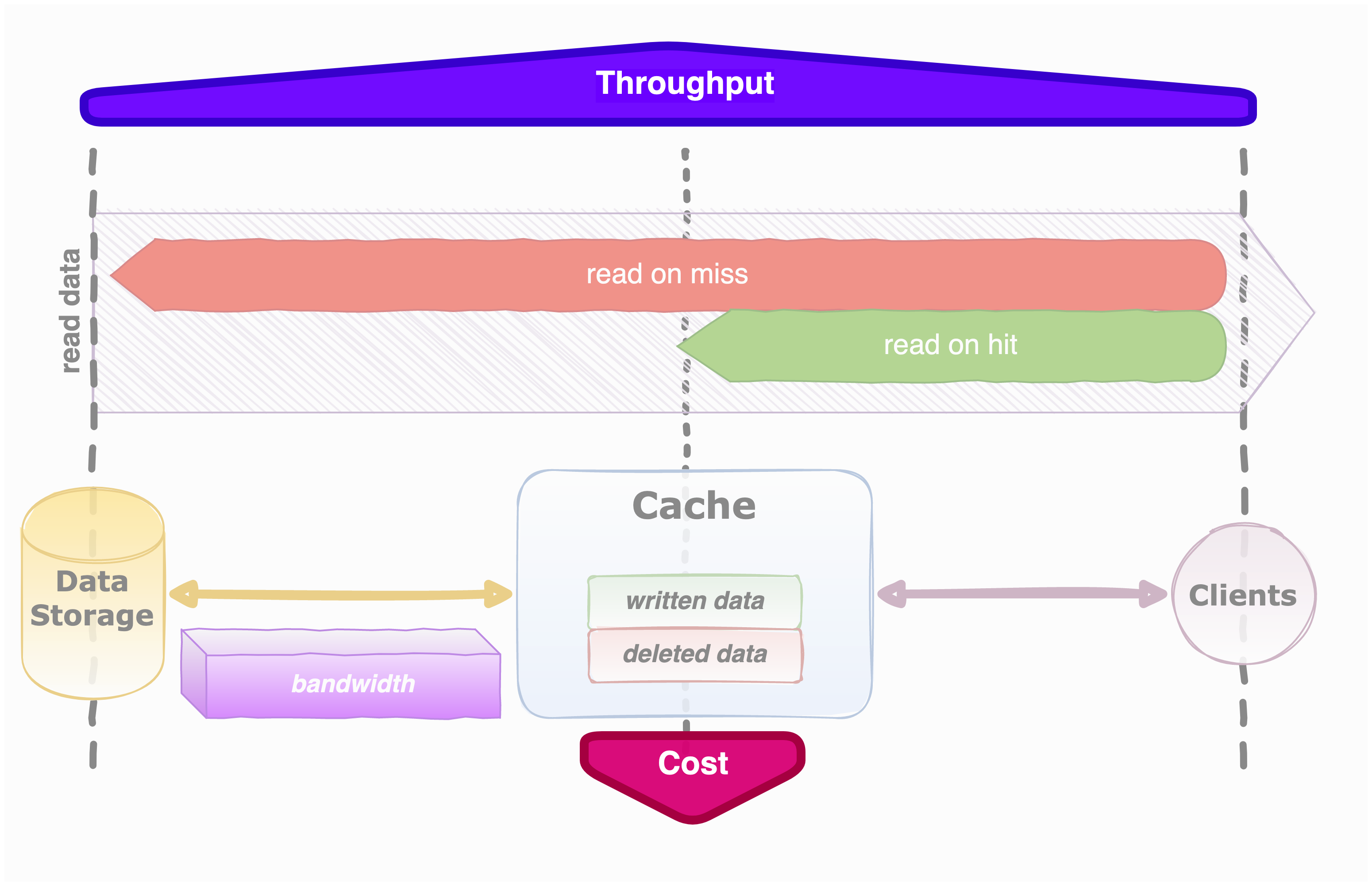}
    \end{center}
    \caption{Projection of the metrics on the simulation environment that shows the several aspects controlled to study the cache behavior}
    \label{fig:cacheEnvironmentMetrics}
\end{figure}

An evident desirable outcome is that the \textit{Throughput} is higher than the \textit{Cost}, because the target is to maximize the cache memory content given a small operational cost. Consequently, we decide to use a \textit{Score} measurement defined as follows:
\begin{equation}
    Score = \frac{TP}{Cost}
    \label{eq:score}
\end{equation}

in order to quantify how much the cache is working in terms of cache operations. The \textit{Score} metric penalizes the cache if it does not serve files from the memory. On the other hand, if most of the content is served from the cache memory, there will be a benefit coming from lower costs and the \textit{Score} ratio will be highly influenced by the \textit{Throughput}.

\section{Experimental Results}
\label{sec:experimentalResults}

All the algorithms have been tested with the data described in~\Cref{sec:datasets}. 
Different cache sizes have been simulated: 100 TiB, 200 TiB, 500 TiB, 1000 TiB. 

\begin{figure}
\centering
\includegraphics[width=.45\textwidth]{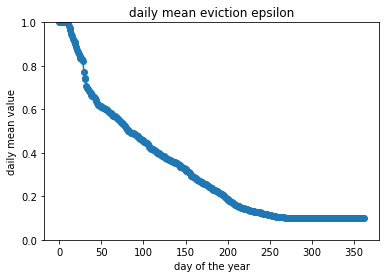}
\includegraphics[width=.45\textwidth]{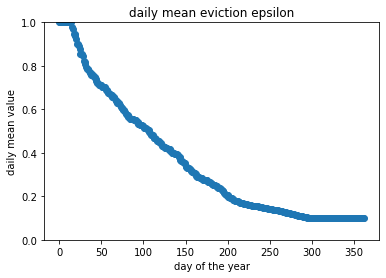}
\caption{DQN QCache $100$ TiB simulation: mean $\epsilon$ daily value as a function of the day of the year for the Addition Agent and the Eviction Agent, respectively.}
\label{fig:epsilon}
\end{figure}

In the DQN QCache algorithm, the $\epsilon$ decay rate is set in order to make sure that the first part of the year is dominated by exploration (i.e., higher $\epsilon$ values were considered), while in the second one the algorithm tends to exploit the gained knowledge (i.e, lower $\epsilon$ values were considered).
This behavior is shown in \Cref{fig:epsilon} where the daily mean value of $\epsilon$ as a function of the day of the year, both for the addition and eviction agents, is reported. For the SCDL algorithms, the $\epsilon$ decay rate is much higher.

Moreover, in DQN QCache approach, both DNNs are 2-hidden-layer (using sigmoid activation) feed-forward networks with 2 output nodes (using linear activation), implemented with Adam optimizer (with 0.001 as learning rate) and Huber loss function. $h_{window}$ is set to 100000 for Addition agent, and to 200000 for the Eviction agent. 

We compared the results obtained with the aforementioned algorithms SCDL, SDCL2 (implemented with different eviction approaches: simple LRU, eviction when memory full, eviction at the end of the day, eviction every K requests) and DQN QCache (implemented with different values of eviction frequency), with the results achieved with a "write everything" approach implemented with different eviction algorithms (LRU, LFU, Biggest Files first, Smallest Files first), since the latter are the most used in caching environments.  Results are shown in Table \ref{tab:results},

\begin{table}[htbp]
\scriptsize
\centering
\begin{tabular}{ p{5.5cm}||p{2cm}|p{2cm}|p{2cm} }
 \textbf{Algorithm - 100 TiB} & \textbf{Score} & \textbf{Throughput} & \textbf{Cost}\\
\textit{DQN} (k50000) & $\pmb{0.34}$ & $0.40$ & $\pmb{1.19}$ \\
SCDL & $0.26$ & $0.45$ & $1.74$ \\
SCDL2 - noEviction  & $0.25$ & $0.45$ & $1.82$ \\
SCDL2 - onFree & $0.22$ & $0.41$ & $1.87$ \\
SCDL2 - onDayEnd & $0.20$ & $0.39$ & $1.93$ \\
Write everything + LRU & $0.19$ & $\pmb{0.50}$ & $2.66$ \\
SCDL2 - onK & $0.17$ & $0.34$ & $2.04$ \\
Write everything + LFU & $0.15$ & $0.43$ & $2.86$ \\
Write everything + Size Big & $0.12$ & $0.37$ & $3.05$ \\
Write everything + Size Small & $0.11$ & $0.36$ &  $3.09$ \\
 \toprule
 \textbf{Algorithm - 200 TiB} & \textbf{Score} & \textbf{Throughput} & \textbf{Cost}\\
\textit{DQN} (k100000) & $\pmb{0.41}$ & $0.45$ & $\pmb{1.12}$ \\
\textit{DQN} (k50000) & $0.35$ & $0.41$ & $1.16$ \\
SCDL  & $0.33$ & $0.55$ & $1.65$ \\
SCDL2 - noEviction  & $0.32$ & $0.54$ & $1.65$ \\
SCDL2 - onFree & $0.28$ & $0.48$ & $1.73$ \\
SCDL2 - onDayEnd & $0.25$ & $0.45$ & $1.83$ \\
Write everything + LRU & $0.24$ & $\pmb{0.59}$ & $2.40$ \\
Write everything + LFU & $0.20$ & $0.52$ & $2.58$ \\
SCDL2 - onK & $0.17$ & $0.35$ & $2.04$ \\
Write everything + Size Big & $0.15$ & $0.42$ & $2.89$ \\
Write everything + Size Small & $0.13$ & $0.39$ & $2.98$ \\
 \toprule
 \textbf{Algorithm - 500 TiB} & \textbf{Score} & \textbf{Throughput} & \textbf{Cost}\\
\textit{DQN} (k250000) & $\pmb{0.54}$ & $0.53$ & $\pmb{0.99}$ \\
SCDL2 - noEviction  & $0.53$ & $0.69$ & $1.30$ \\
SCDL & $0.51$ & $0.72$ & $1.41$ \\
SCDL2 - onFree & $0.39$ & $0.60$ & $1.55$ \\
Write everything + LRU & $0.39$ & $\pmb{0.74}$ & $1.90$ \\
Write everything + LFU & $0.32$ & $0.67$ & $2.11$ \\
\textit{DQN} (k50000) & $0.31$ & $0.47$ & $1.51$ \\
SCDL2 - onDayEnd & $0.28$ & $0.49$ & $1.75$ \\
Write everything + Size Big & $0.22$ & $0.54$ & $2.52$ \\
Write everything + Size Small & $0.18$ & $0.48$ & $2.70$ \\
SCDL2 - onk & $0.17$ & $0.36$ & $2.04$ \\
 \toprule
 \textbf{Algorithm - 1000 TiB} & \textbf{Score} & \textbf{Throughput} & \textbf{Cost}\\
SCDL2 - noEviction  & $\pmb{0.78}$ & $0.80$ & $1.01$ \\
SCDL & $0.71$ & $0.83$ & $1.17$ \\
\textit{DQN} (k500000)& $0.69$ & $0.60$ & $\pmb{0.86}$ \\
Write everything + LRU & $0.59$ & $\pmb{0.86}$ & $1.45$ \\
SCDL2 - onFree & $0.55$ & $0.70$ & $1.27$ \\
Write everything + LFU & $0.48$ & $0.80$ & $1.65$ \\
\textit{DQN} (k50000)& $0.34$ & $0.41$ & $1.21$ \\
Write everything + Size Big & $0.33$ & $0.68$ & $2.07$ \\
SCDL2 - onDayEnd & $0.30$ & $0.53$ & $1.78$ \\
Write everything + Size Small & $0.26$ & $0.59$ & $2.32$ \\
SCDL2 - onk & $0.18$ & $0.36$ & $2.04$ \\

\end{tabular}
\caption{Comparison of results of different algorithms (daily values averaged across the year): SCDL, described in section \ref{sec:SCDL}; SCDL2, described in section \ref{sec:SCDL2}, implemented with different eviction policies: simple LRU (noEviction), eviction when memory full (onFree), eviction at the end of the day (onDayEnd), eviction every K requests (onk) where $K=8192$; DQN, described in section \ref{sec:DQN}, implemented with different eviction frequencies (indicated as kN, where N is the frequency); Write everything approaches with different eviction policies: Least Recently Used (LRU), Least Frequently Used (LFU), delete biggest files first (Size Big) and delete smallest files first (Size Small). The best result for each metric is displayed in bold.}
\label{tab:results}
\end{table}

Looking at the results reported in \Cref{tab:results} we can observe that the approaches we are proposing show overall better performances, in terms of \textit{Score}, compared to the standard cache policies, like the LRU and LFU deleting systems. 
However, the LRU method always reached the best \textit{Throughput} value, also for higher cache sizes. But, considering the \textit{Score} also the \textit{Cost} of the operations, LRU cannot outstand with respect to the proposed RL approaches.

Furthermore, we can attest that the different mechanisms used to free the cache memory content have a deeper impact on the general caching performances with respect to the simple file-filtering. Despite that, in the cases of larger caches, the best algorithms use only the Addition Agent, that means the strategy for file eviction surely needs to be improved and further optimized. 
Moreover, it is clear that standard policies in general cannot compete with both SCDL2 and DQN QCache in terms of Cost (see ~\Cref{eq:cost}). Indeed our RL approaches generally make the cache less active by doing the minimum number of operations to maintain a good cache composition. This results in a lower amount of written and deleted files. Although, the presence of missed files may still affect the network. 

To summarize, the two agents affect the cache environment differently. The Addition Agent is the main responsible for reducing the amount of written data and selecting files to store in a more rigorous way. Hence, the available network bandwidth is more used and the \textit{Throughput} is increased. The Eviction Agent affects the presence of the files in the cache. Hence, its main outcome is to increase the \textit{Throughput} and to decrease the \textit{Cost}, maintaining a higher level of \textit{Read on Hit}.

Finally, and most importantly, it is crucial to underline a key aspect: in the present simulation each delete or write operation is considered to be timeless. Reason why, we are expecting that, in a real-world scenario, a cache system that is less busy in writing and removing files will be surely readier to distribute the requested files to the clients, i.e, it will be more efficient and will provide a final better use experience. Indeed the RL approaches are always the top ranked with respect to the \textit{Score}, as clear evidence of this fact.

\section{Conclusions}
\label{sec:conclusions}

Recently the CMS community at CERN has started to experiment with new models to manage the whole computing infrastructure due to upcoming updates and the huge amount of data foreseen for the next years, exploring the possibility of 
moving towards a Data Lake model. This new scenario imposes to find more effective solutions to the data caching problem. Thus, the role of the cache becomes a key to effective and efficient data access.

In this work, we introduced three different RL caching algorithms. We defined the metrics to compare them  to standard caching policies. We performed a set of tests using different cache sizes and a real-world dataset based on historical monitoring data about CMS experiment analysis jobs. We were also able to obtain a first feedback on the effectiveness of these approaches.

We can conclude that the RL caching algorithms we implemented showed better overall performances in terms of \textit{Score}, and especially in terms of \textit{Cost}, with respect to the standard policies using, for example, LRU eviction strategy. Our RL approaches make the cache less active by doing a lower number of operations to maintain a good cache composition. This results in a lower amount of written and deleted data. While the presence of missed files still affects the network, we want to underline that we are expecting that in a real-world scenario (where the time domain is taken into account), a cache system that is less busy in writing and removing files will be surely more responsive and quicker to serve the requested files to the clients.

\section{Acknowledgments}
\label{sec:ack}
The authors thank the CMS collaboration, and in particular the Machine Learning and Offline Software and Computing groups for the valuable discussions that helped the development of this work.
  
\bibliography{references} 

\begin{thebibliography}{10}

\bibitem{Pettersson:291782}
T.~S. Pettersson and P.~Lefèvre, ``{\relax The Large Hadron Collider:
  conceptual design},'' tech. rep., Oct 1995.

\bibitem{aad2008atlas}
{\relax The ATLAS Collaboration}, ``{\relax The ATLAS experiment at the CERN
  Large Hadron Collider},'' {\em Journal of Instrumentation}, vol.~3,
  p.~S08003, 2008.

\bibitem{collaboration2008cms}
{\relax The CMS Collaboration}, ``The {CMS} experiment at the {CERN} {LHC},''
  {\em Journal of Instrumentation}, vol.~3, pp.~S08004--S08004, aug 2008.

\bibitem{aamodt2008alice}
{\relax The ALICE Collaboration}, ``{\relax The ALICE experiment at the CERN
  LHC},'' {\em Journal of Instrumentation}, vol.~3, no.~08, p.~S08002, 2008.

\bibitem{alves2008lhcb}
{\relax The LHCb Collaboration}, ``{\relax The LHCb detector at the LHC},''
  {\em Journal of instrumentation}, vol.~3, no.~08, p.~S08005, 2008.

\bibitem{offcompupdate}
{\relax CMS Offline Software and Computing}, ``{\relax CMS Phase-2 Computing
  Model: Update Document},'' 2022.
\newblock CERN-CMS-NOTE-2022-008, available on the CERN Document Server as
  \url{https://cds.cern.ch/record/2815292}.

\bibitem{bird2019architecture}
I.~Bird, S.~Campana, M.~Girone, X.~Espinal, G.~McCance, and J.~Schovancov{\'a},
  ``{\relax Architecture and prototype of a WLCG data lake for HL-LHC},'' in
  {\em EPJ Web of Conferences}, vol.~214, p.~04024, EDP Sciences, 2019.

\bibitem{kadochnikov2018wlcg}
I.~Kadochnikov, I.~Bird, G.~McCance, J.~Schovancova, M.~Girone, S.~Campana, and
  X.~E. Currul, ``{\relax WLCG data lake prototype for HL-LHC},'' {\em Advisory
  committee}, p.~127, 2018.

\bibitem{dataLakeDixon}
J.~Dixon, ``Pentaho, hadoop and data lakes.''
  \url{https://jamesdixon.wordpress.com/2010/10/14/pentaho-hadoop-and-data-lakes/},
  2010.
\newblock Last check April 9, 2020.

\bibitem{adhikari2012unreeling}
V.~K. Adhikari, Y.~Guo, F.~Hao, M.~Varvello, V.~Hilt, M.~Steiner, and Z.-L.
  Zhang, ``\relax{Unreeling netflix: Understanding and improving multi-CDN
  movie delivery},'' in {\em 2012 Proceedings IEEE INFOCOM}, pp.~1620--1628,
  IEEE, 2012.

\bibitem{sutton2018reinforcement}
R.~S. Sutton and A.~G. Barto, {\em Reinforcement learning: An introduction}.
\newblock MIT press, 2018.

\bibitem{mnih2013playing}
V.~Mnih, K.~Kavukcuoglu, D.~Silver, A.~Graves, I.~Antonoglou, D.~Wierstra, and
  M.~Riedmiller, ``Playing atari with deep reinforcement learning,'' {\em arXiv
  preprint arXiv:1312.5602}, 2013.

\bibitem{mnih2015human}
V.~Mnih, K.~Kavukcuoglu, D.~Silver, A.~A. Rusu, J.~Veness, M.~G. Bellemare,
  A.~Graves, M.~Riedmiller, A.~K. Fidjeland, G.~Ostrovski, S.~Petersen,
  C.~Beattie, A.~Sadik, I.~Antonoglou, H.~King, D.~Kumaran, D.~Wierstra,
  S.~Legg, and D.~Hassabis, ``{Human-level control through deep reinforcement
  learning},'' {\em Nature}, vol.~518, no.~7540, pp.~529--533, 2015.

\bibitem{podlipnig2003survey}
S.~Podlipnig and L.~B{\"o}sz{\"o}rmenyi, ``A survey of web cache replacement
  strategies,'' {\em ACM Computing Surveys (CSUR)}, vol.~35, no.~4,
  pp.~374--398, 2003.

\bibitem{lei2017deep}
L.~Lei, L.~You, G.~Dai, T.~X. Vu, D.~Yuan, and S.~Chatzinotas, ``A deep
  learning approach for optimizing content delivering in cache-enabled
  hetnet,'' in {\em 2017 international symposium on wireless communication
  systems (ISWCS)}, pp.~449--453, IEEE, 2017.

\bibitem{narayanan2018deepcache}
A.~Narayanan, S.~Verma, E.~Ramadan, P.~Babaie, and Z.-L. Zhang, ``Deepcache: A
  deep learning based framework for content caching,'' in {\em Proceedings of
  the 2018 Workshop on Network Meets AI \& ML}, pp.~48--53, 2018.

\bibitem{lykouris2018competitive}
T.~Lykouris and S.~Vassilvitskii, ``{\relax Competitive caching with machine
  learned advice},'' {\em arXiv preprint arXiv:1802.05399}, 2018.

\bibitem{herodotou2019autocache}
H.~Herodotou, ``Autocache: Employing machine learning to automate caching in
  distributed file systems,'' {\em International Conference on Data Engineering
  Workshops (ICDEW)}, pp.~133--139, 2019.

\bibitem{sadeghi2019deep}
A.~Sadeghi, G.~Wang, and G.~B. Giannakis, ``Deep reinforcement learning for
  adaptive caching in hierarchical content delivery networks,'' {\em IEEE
  Transactions on Cognitive Communications and Networking}, vol.~5, no.~4,
  pp.~1024--1033, 2019.

\bibitem{dulac2015deep}
G.~Dulac-Arnold, R.~Evans, H.~van Hasselt, P.~Sunehag, T.~Lillicrap, J.~Hunt,
  T.~Mann, T.~Weber, T.~Degris, and B.~Coppin, ``{\relax Deep reinforcement
  learning in large discrete action spaces},'' {\em arXiv preprint
  arXiv:1512.07679}, 2015.

\bibitem{zhong2018deep}
C.~Zhong, M.~C. Gursoy, and S.~Velipasalar, ``A deep reinforcement
  learning-based framework for content caching,'' in {\em 2018 52nd Annual
  Conference on Information Sciences and Systems (CISS)}, pp.~1--6, IEEE, 2018.

\bibitem{alabed}
S.~Alabed, ``\relax{RLCache: Automated Cache Management Using Reinforcement
  Learning},'' {\em arXiv preprint arXiv:1909.13839}, 2019.

\bibitem{scdl_code}
M.~Tracolli, ``Open source code,'' 2022.
\newblock Available at
  \url{https://github.com/Cloud-PG/smart-cache/tree/master}.

\bibitem{SCDL}
M.~Tracolli, M.~Baioletti, V.~Poggioni, and D.~Spiga, ``{\relax Caching
  suggestions using Reinforcement Learning },'' tech. rep., University of
  Perugia, Dept. Mathematics and Computer Science, 2020.

\bibitem{dql_code}
T.~Tedeschi, ``Open source code,'' 2022.
\newblock Available at
  \url{https://github.com/Cloud-PG/smart-cache/tree/dQl_add_evic_no_gym}.

\bibitem{kuznetsov2016predicting}
V.~Kuznetsov, T.~Li, L.~Giommi, D.~Bonacorsi, and T.~Wildish,
  ``\relax{Predicting dataset popularity for the CMS experiment},'' {\em arXiv
  preprint arXiv:1602.07226}, 2016.

\bibitem{meoni2018dataset}
M.~Meoni, R.~Perego, and N.~Tonellotto, ``\relax{Dataset popularity prediction
  for caching of CMS big data},'' {\em Journal of Grid Computing}, vol.~16,
  no.~2, pp.~211--228, 2018.

\end{thebibliography}
\bibliographystyle{ieeetr}

\end{document}